\documentclass[usenatbib]{mnras}

\usepackage{graphicx}	
\usepackage{amsmath}	
\usepackage{amssymb}	
\graphicspath{{}{./}}
\DeclareGraphicsExtensions{.pdf,.png,.jpg}
\newcommand{\vct}[1]{\mbox{\boldmath $#1$}}

\title[Planetary migration]{A new and simple prescription for planet orbital migration and eccentricity damping by planet-disc interactions based on dynamical friction}

\author[S. Ida et al.]{
Shigeru Ida$^{1}$,\thanks{E-mail: ida@elsi.jp}
Takayuki Muto$^{3}$,
Soko Matsumura$^{2}$,
and 
Ramon Brasser$^{1}$
\\
$^{1}$Earth Life Science Institute, Tokyo Institute of Technology, Tokyo 152-8550, Japan\\
$^{2}$Division of Physics, University of Dundee, Dundee, DD1 4HN, UK\\
$^{3}$Division of Liberal Arts, Kogakuin University, Tokyo,192-0015, Japan
}

\begin{document}
\label{firstpage}
\pagerange{\pageref{firstpage}--\pageref{lastpage}}
\maketitle

\begin{abstract}
{During planet formation gravitational interaction between a planetary embryo and the protoplanetary gas disc causes 
orbital migration of the planetary embryo, which plays an important role in shaping the final planetary system.
While migration sometimes occurs in the supersonic regime, wherein the relative velocity between
the planetary embryo and the gas is higher than the sound speed, migration prescriptions proposed thus far describing 
the planet-disc interaction force and the timescales of orbital change in the supersonic regime
are inconsistent with one another. Here we discuss the details of existing prescriptions in the literature and 
derive a new simple and intuitive formulation for planet-disc interactions based on dynamical friction that can be
applied in both supersonic and subsonic cases. While the existing prescriptions assume particular disc models,
ours include the explicit dependence on the disc parameters; hence it can be applied to discs with any radial 
surface density and temperature dependence (except for the local variations with radial scales less than the disc scale height). Our prescription will reduce the uncertainty originating from different literature 
formulations of planet migration and will be an important tool to study planet accretion processes, 
especially when studying the formation of close-in low-mass planets that are commonly found in exoplanetary systems.} 
\end{abstract}




\section{Introduction}
Kepler transit surveys and ground-based radial velocity surveys
have discovered hundreds of exoplanetary systems that have multiple close-in low-mass planets.
One formation scenario for these systems is through the local accretion of planetary embryos in the systems' 
outer regions followed by their inward migration and late mutual collisions
\citep[e.g.][]{Ogihara09, Ida10, Morbidelli16}.
During these processes, the orbital eccentricities of the planetary embryos are pumped up.  
As we will show, when the eccentricity $e$
exceeds the disc gas aspect ratio $h$, which is as small as $\sim 0.02{\rm -}0.03$
in the proximity of the host star, the planet-disc interactions are often in the supersonic regime.

For a planetary embryo embedded in the disc gas,
the planet-disc interactions always damp the orbital eccentricity 
(Tanaka \& Ward 2004, hereafter referred to as \citet{Tanaka04}).
The semimajor axis generally decreases for a circular orbit 
in the isothermal disc case, which is called ``type I migration" \citep[e.g.][]{Tanaka02}. 
The literature is filled with analytical and fitting formulae for the eccentricity damping and rate of migration as a function of eccentricity 
(Papaloizou \& Larwood 2000, hereafter referred to as \citet{Papaloizou00};
\citet{Tanaka04}; Muto et al. 2011, hereafter referred to as \citet{Muto11}:
Cresswel \& Nelson 2008, hereafter referred to as \citet{Cresswell08};
\citet{Coleman14}).
The formulae are complicated and have different forms from one another. We review these in section 2.

N-body simulations are often used to study the formation of close-in low-mass planets,
taking the effects of the planet-disc interactions into account. 
Because the migration and coagulation timescales are much longer than available timescales by hydrodynamical simulations, coupled N-body and hydrodynamical 
simulations are not realistic.   
Instead, the N-body simulations have included these effects 
by addition of the force term of the interactions to the equations of motion,
according to one of the several established prescriptions listed above
\citep[e.g.][]{Terquem07, Ogihara09, Coleman14, Cossou14, Matsumura17, Izidoro19}.

Some formulae clearly show that the net torque from the disc changes sign 
near the subsonic-supersonic boundary
\citep{Papaloizou00, Cresswell08, Coleman14}.
Several works have called the orbital angular momentum damping timescale due to the disc torque as the 
``migration timescale," so that the sign change looks to imply that outward migration occurs in the supersonic case.
However, the torque reversal does not necessarily mean outward migration actually happens in the supersonic case, 
as we will discuss in section 2 \citep[also see,][]{Baruteau14,Grishin15,Matsumura17}.
\citet{Brasser18} demonstrated through N-body simulations that the resonant trapping and 
collisions between the planetary embryos near the disc inner edge sensitively depend on the adopted migration and damping
prescription. 
Because $e$ is excited to a value near the subsonic-supersonic boundary
due to the resonances,
the choice of the prescriptions
affects whether the innermost planetary embryos are pushed into 
the inner cavity or not, and whether the resonant chain remains stable.
Thus, the uncertainty and confusion in the planet-disc interaction formulae
across the subsonic and supersonic regimes 
results in uncertainty in the results of N-body simulations.

In this paper, we discuss the formulae of planet-disc interactions that cover both subsonic and supersonic cases.
We further propose a new prescription for type I migration based on dynamical friction, and we make clear what 
should be investigated in the prescription to further understand the formation and migration of multiple 
close-in planets in exoplanetary systems. 

\section{Formulae of Eccentricity Damping and Migration in Previous Studies}

In this section, we summarize the existing formulae of planet-disc interactions for finite eccentricity orbits.
While we also consider the supersonic case, $e \ga h$, we assume that $e^2 \ll 1$.
In this paper, we also assume $i \ll e$ to
neglect the inclination damping in most of the parts for simplicity,
although we consider 3D interactions.
We briefly comment on the inclination damping in Appendix D.

\subsection{Supersonic/subsonic interactions}

We first make clear the relation between the orbital eccentricity of a planetary embryo and 
supersonic/subsonic interactions.

The relative velocity between the disc gas and a planetary embryo in an orbit with the eccentricity $e$ 
is approximately $\Delta v \sim e v_{K} = e r \Omega_{K}$, where $v_{K}$ and $\Omega_{K}$ are the 
Keplerian velocity and frequency at the radius $r$ from the host star. 
The sound velocity is given by $c_s = H \Omega_{K}$,
where $H$ is the disc scale height.
Therefore, $e \sim h \equiv H/r$ is equivalent to $\Delta v \sim c_s$, 
The supersonic interaction, defined by $\Delta v > c_s$, is equivalent to $e >h$.

Because the aspect ratio is usually as small as $h \sim 0.02{\rm -}0.03$
in inner regions of the disc, supersonic interactions are easily realized.
During orbital migration, the planetary embryos are often trapped in mutual mean-motion resonances
to form a resonant chain, in particular, near the disc inner edge at which 
the innermost planetary embryo's migration significantly slows down 
\citep[e.g.][]{Ogihara09, Matsumura17}.
In this resonant chain, resonant perturbations excite 
$e$, while the planet-disc interactions damp $e$.
The equilibrium value is $e \sim h$ \citep{Goldreich14}.
Secular perturbations from giant planet(s) in outer disc regions may also pump up
$e$ over $0.02{\rm -}0.03$.  
During close scattering between the planetary embryos,
their eccentricities are excited up to $e_{\rm esc} \sim v_{\rm esc}/v_{\rm K}$
where $v_{\rm esc}$ is the surface escape velocity of the planetary embryos.
For Earth-mass planetary embryos at $\sim 0.1\, {\rm au}$
around a solar-type star, $e_{\rm esc} \sim 0.1$.
After the scattering, $e$ must be in the supersonic regime, although 
it is damped by the planet-disc interaction afterward.

We need to be careful about the case of $e < h$.
In this case, disc shear motion determines $\Delta v$
and the torque density is the largest for shear flow at the semimajor axis difference $\Delta a \sim H$ \cite[e.g.][]{Artymowicz93,Ward97,Miyoshi99}.
While the dominant interaction is not caused by close encounters, 
the relative velocity in that case is given by the shear velocity, 
$\Delta v \sim (3/2) \Delta a \Omega_{K} \sim (3/2) H \Omega_{K} \sim 3 c_s/2$,
which is near the supersonic/subsonic boundary. 

Although the complexity exists in the case of $e < h$,
in this paper, we call the cases of $e<h$ and $e>h$ as ``subsonic" and ``supersonic" cases,
respectively.

\subsection{Migration, Eccentricity Damping, and Angular Momentum Transfer}

The specific angular momentum of a planetary embryo in an orbit with eccentricity $e$ and semi-major axis $a$ is given by
\begin{equation}
\ell = \sqrt{GM_* a(1-e^2)},
\label{eq:ell}
\end{equation}
where $G$ is the gravitational constant and $M_*$ is the host star mass.
The rate of change of $\ell$ is split into two parts:
\begin{equation}
\frac{1}{\ell}\frac{d\ell}{dt}
= \frac{1}{2a}\frac{da}{dt} - \frac{e}{1-e^2}\frac{de}{dt}.
\label{eq:ldot}
\end{equation}
In the past papers of the linear analysis, the ``type I migration rate" often referred to $(1/\ell)(d\ell/dt)$ but not $(1/a)(da/dt)$.
However, $(1/\ell)(d\ell/dt)$ represents orbital migration 
only for zero eccentricity.
Equation~(\ref{eq:ldot}) can be rewritten as
\begin{equation}
\tau_m^{-1} = \frac{1}{2}\tau_a^{-1} - \frac{e^2}{1-e^2}\tau_e^{-1},
\label{eq:ldot2}
\end{equation}
where we define the rate of decrease (inverse of timescales) of angular momentum, semimajor axis,
and eccentricity as
\begin{equation}
\tau_m^{-1} = -\frac{d\ell/dt}{\ell}, \;\; 
\tau_a^{-1} = -\frac{da/dt}{a}, \;\;
\tau_e^{-1} = -\frac{de/dt}{e}.
\end{equation}
With these definitions, migration is inward when $\tau_a >0$, the eccentricity is damped when $\tau_e>0$
and the planetary embryo loses angular momentum when $\tau_m>0$. 
When $e \ll h$, $\tau_a^{-1} \simeq 2\tau_m^{-1}$
and $\tau_m^{-1}$ is identified as the semimajor axis change rate (``migration rate")
except for the factor of 2.

For $e \ll h$ in a locally isothermal disc we have
$\tau_a^{-1} \sim h^2 \tau_e^{-1}$ \citep[e.g.][]{Tanaka02, Tanaka04}.
With this relation, Eq.~(\ref{eq:ldot2}) can be rewritten as
\begin{equation}
\tau_m^{-1} \simeq \frac{1}{2} \tau_a^{-1} - e^2 \tau_e^{-1}
\sim (h^2 - e^2) \tau_e^{-1}.
\label{eq:tau_m}
\end{equation}
As we will show, in the supersonic regime ($e > h$), both eccentricity and semimajor axis are damped, 
when the planetary embryo is embedded in the disc, i.e. $\tau_e^{-1} > 0$ and $\tau_a^{-1} > 0$,
except for \citet{Papaloizou00}'s formulae.
Although Eq.~(\ref{eq:tau_m}) is guaranteed only when $e \ll h$,
it suggests that $\tau_m^{-1}$ changes sign when $e\sim h$. We will show that the sign change actually occurs at $e\sim h$ with
all the proposed formulae. 

\subsection{Various Formulae in Previous Studies}

\citet{Goldreich79} derived the ``migration" rate 
of a planetary embryo through linear calculation. 
\citet{Tanaka02} performed a more detailed analysis taking both the corotation and Lindblad resonances into account, and calculated the ``migration" rate as
\begin{equation}
\tau_{m,e=0}^{-1} = \frac{C_{\rm T}}{2} h^2 t_{\rm wave}^{-1},
\end{equation}
where
\begin{align}
& t_{\rm wave}^{-1} = \left(\frac{M_p}{M_*}\right)\left(\frac{\Sigma r^2}{M_*}\right)
h^{-4}  \Omega_K, \label{eq:twave}\\
& C_{\rm T} = (2.73+1.08\,p+0.87\,q),  \label{eq:facC}
\end{align}
and where $\Sigma$ is the disc surface density, $M_*$ is the host star mass, $p = - d\ln \Sigma/d\ln r$, and 
we also included temperature radial dependence $q = - d\ln T/d\ln r$ through pressure ($P$) and scale height ($H$).
Both \citet{Goldreich79} and \citet{Tanaka02} assumed circular orbits, so that the semimajor axis decrease rate is 
$\tau_a^{-1} = 2\tau_m^{-1}$. 

When we consider eccentric orbits, 
the radial position of the planetary embryo, $r$, changes during one orbit and  
$t_{\rm wave}^{-1}$ also changes with time.
Hereafter in this paper, we denote the instantaneous value at $r$ as
$t_{{\rm wave},r}^{-1}$ and the value at the planetary embryo's semimajor axis 
as $t_{\rm wave}^{-1}$ without the subscript ``$r$."

\citet{Tanaka04} derived the eccentricity damping timescale and
the (specific) damping forces in the locally isothermal disc through a 3D linear calculation,
 \begin{equation}
\tau_e^{-1}  = 0.780 \, t_{\rm wave}^{-1}, 
 \end{equation}
and
 \begin{align}
 & \vct{F}_{\rm damp} = \frac{f_{r,\rm TW}}{t_{\rm wave}}\vct{e}_r +
 \frac{f_{\theta,\rm TW}}{t_{\rm wave}}\vct{e}_\theta. \label{eq:TW} \\
  & f_{r,\rm TW} = 0.114 (v_\theta - r\Omega_K) + 0.176\,  v_r,  \label{eq:TWr}\\
 &  f_{\theta,\rm TW} = -1.736 (v_\theta - r\Omega_K) + 0.325\; v_r, \label{eq:TWtheta}
\end{align} 
where $\vct{e}_r$ and $\vct{e}_\theta$ are the unit vectors in radial and azimuthal directions.  
In their derivations, $e < h$ was assumed, and they included the effects from density spiral waves that develop
only for $e < h$ (see the discussion in section 3.1).
Because they considered the case of $e < h$, they identified $t_{{\rm wave},r}^{-1}$
as $t_{\rm wave}^{-1}$.

\citet{Papaloizou00} derived $\tau_e^{-1}$ and $\tau_m^{-1}$ 
for eccentric orbits
in a 2D disc model, using a similar Fourier expansion as \citet{Goldreich79}. The 3D effect was considered by 
incorporating a softening parameter ($\epsilon$) for the planetary embryo's gravitational potential. They considered the temperature variation of 
$q = 1$ so that $h = c_s/v_K$ is constant with $r$ ($c_s \propto \sqrt{T} \propto r^{-q/2}$),
and neglected the corotation torque by assuming that $p=3/2$. 
(In \citet{Tanaka02} and \citet{Tanaka04}, both $e \ll h$ and a locally constant
$T$ were assumed, while they allowed any radial power-law dependence of 
$\Sigma$, $P$, and $H$). 
Although their disc parameters were fixed,
\citet{Papaloizou00} were the first to derive a formulation of $\tau_e^{-1}$ and $\tau_m^{-1}$ that is applicable in both the subsonic and the supersonic cases. 
Because the supersonic correction factors are given by functions of $e/h$,
we define $\hat{e}=e/h$.
\citet{Papaloizou00}'s results are given by (see Appendix A)
\begin{align}
\tau_e^{-1} & \simeq  4.26 \, \left(\frac{\epsilon}{0.5H}\right)^{-2.5} \left(1 + \frac{1}{4} \hat{e}^{3} \right)^{-1} \, t_{\rm wave}^{-1},
\label{eq:PL00_te} \\
\tau_m^{-1} & \simeq 7.33 \, \left(\frac{\epsilon}{0.5H}\right)^{-1.75} h^{2}  
\frac{1 - \left( \frac{\hat{e}}{1.1} \right)^4} 
{1 + \left( \frac{\hat{e}}{1.3} \right)^5} \; t_{\rm wave}^{-1},
\label{eq:PL00_tm}
\end{align}
where $\epsilon \sim (0.4{\rm -}1) \,H$ is often adopted to mimic 3D results.
As shown in Appendix A, we used $M_{\rm GD} = \int^r 2\pi r \Sigma dr = 4\pi \Sigma r^2$ in the original formulae by \citet{Papaloizou00},
considering their model with $\Sigma \propto r^{-3/2}$,
although \citet{Papaloizou00} did not provide an explicit expression of $M_{\rm GD}$.
If a simpler expression, $M_{\rm GD} \sim \pi \Sigma r^2$ is used,
the above rates of changes of \citet{Papaloizou00} decrease by a factor of 4. 

\begin{figure}\begin{center} 
\includegraphics[width=\columnwidth]{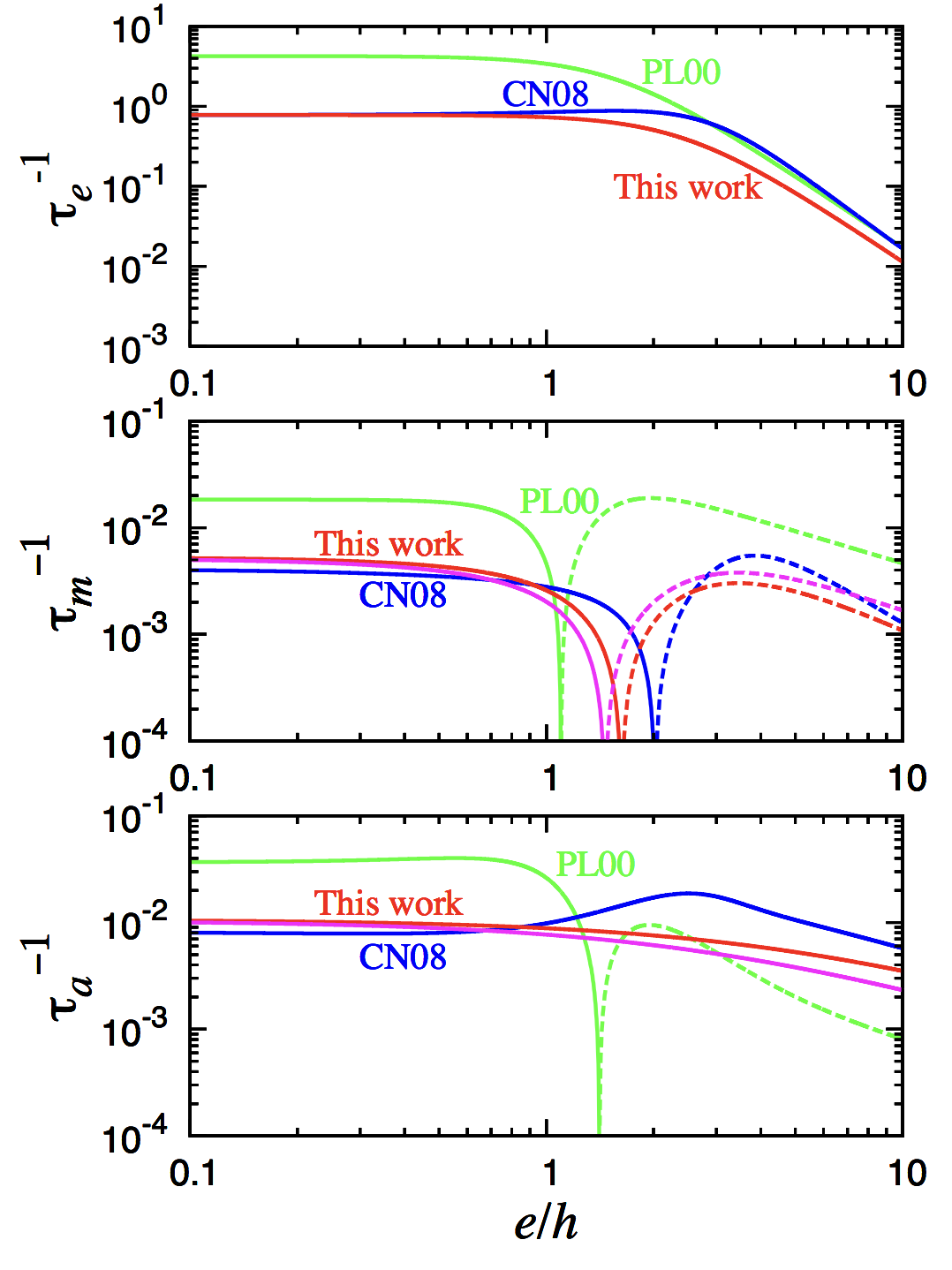}
\end{center}
\caption{
The $e$-damping rate ($\tau_e^{-1}$),
the angular momentum transfer rate ($\tau_m^{-1}$),
and the $a$-change rate ($\tau_a^{-1}$) in the unit of $t_{\rm wave}^{-1}$,
predicted by the three different expressions are plotted as functions of $e/h$
for $h=0.05$.
The dashed lines represent negative values.
The green and blue curves are
formulae by \citet{Papaloizou00} (Eqs.~(\ref{eq:PL00_te}) 
and (\ref{eq:PL00_tm})) for $p=1.5$ and $q=1$
and \citet{Cresswell08} (Eqs.~(\ref{eq:CN08_te}) and (\ref{eq:CN08_tm}))
for $p=0.5$ and $q=1$, respectively.
The red and magenta curves represent the simple formulae we derive in this paper
(Eqs.~(\ref{eq:tau_e_DF}), (\ref{eq:tau_a_DF}), (\ref{eq:tau_m_DF}); ``This work")
for the nominal disc parameters $p=1$ and $q=0.5$ and those for
CN08's disc model with $p=0.5$ and $q=1$, respectively.
In the top panel, the red and magenta curves overlap.
The dependences on $p$ and $q$ of our formulae are explicitly shown in 
Eqs.~(\ref{eq:tau_a_DF}), and (\ref{eq:tau_m_DF}).
We adopt the softening parameters, $\epsilon = 0.5 \, H$.
}\label{fig:mig_time}
\end{figure}

\citet{Papaloizou00} also proposed a force formula 
for combined migration and eccentricity damping, 
\begin{equation}
\frac{d\vct{v}}{dt} = - \frac{\vct{v}}{\tau_m} 
-  2\frac{(\vct{v} \cdot \vct{r})\vct{r}}{r^2 \tau_e} = - \frac{\vct{v}}{\tau_m} -  2\frac{v_r}{\tau_e} \vct{e}_r.
\label{eq:dvdtPL00}
\end{equation}
where $\vct{v}$ is the planetary embryo velocity in the inertial frame and the 1st and the 2nd terms represent ``migration" and $e$-damping.
The former reduces the orbital energy and the latter damps the deviation of orbital motion from the circular Keplerian motion.
While the force formula for the eccentricity damping derived by \citet{Tanaka04} has 
both radial and tangential components\footnote{
Note that \citet{Tanaka04} gave a force formula only for eccentricity damping.}, 
the above $e$-damping force,
$-  2(v_r/\tau_e) \vct{e}_r$, has only a radial component.
Because \citet{Papaloizou00} did not give the derivation of the force formula,
we will give a potential derivation in section 4. 

In the formulation of \citet{Papaloizou00} (Eq.~(\ref{eq:PL00_tm})), $\tau_m$ becomes negative when $e > 1.1\,h$, which appears to imply outward migration in the supersonic regime, 
even in the isothermal case. 
Figure~\ref{fig:mig_time} shows Eqs.~(\ref{eq:PL00_te}) and (\ref{eq:PL00_tm})
and $\tau_a^{-1}$ calculated using Eq.~(\ref{eq:ldot2}) by the green curves.
In \citet{Papaloizou00}'s formulation,
$\tau_a^{-1} < 0$ for $e > 1.3 \, h$.
However, this sign change is very delicate.
Equations~(\ref{eq:PL00_te}) and (\ref{eq:PL00_tm}) show that
$\tau_a^{-1}$ for $\tilde{e} \gg1$ (where $e^2 \ll 1$ is still assumed) is
\begin{align}
\tau_a^{-1} & \simeq 2\, \tau_m^{-1} + 2\, e^2 \tau_e^{-1} \nonumber\\
  & \simeq - 37.2 \left[1 - 0.92 \left(\frac{\epsilon}{0.5H}\right)^{-0.75}\right] \nonumber\\
  & \hspace{0.5cm} \times \left(\frac{\epsilon}{0.5H}\right)^{-1.75} h^2 \tilde{e}^{-1} \, t_{\rm wave}^{-1}.
\label{eq:PL00_ta} 
\end{align}
The value of $\epsilon$ has an uncertainty.
With slightly smaller $\epsilon$ than $0.5H$, 
the effect of the $e$-damping
on the angular momentum transfer is more important and 
$\tau_a^{-1}$ can be positive for $e \gg h$.
As mentioned in the above, $\tau_m^{-1} < 0$ does not necessarily imply outward migration in the supersonic case because of the coupled eccentricity damping.
As we will show below, $\tau_a^{-1}$ is always positive, both in \citet{Cresswell08}'s formulae
and our formulae that are proposed in this paper.

\citet{Cresswell08} adopted different finite eccentricity corrections
by fitting the results of their 3D hydrodynamical simulations. Their disc had $p=1/2$ and $q=1$,
while \citet{Papaloizou00} assumed a disc with $p=3/2$ and $q=1$. 
The power index of $q \sim 1$ is appropriate for the disk regions
where the viscous heating dominates, while
$q \sim 0.5$  is appropriate for irradition-dominated regions \citep[e.g.][]{Ida16}.
\citet{Cresswell08}'s fitting formulae are
\begin{align}
& \tau_e^{-1} \simeq 0.78 \,
\left(1 - 0.14 \hat{e}^{2} + 0.06 \hat{e}^{3} \right)^{-1} t_{\rm wave}^{-1}, \label{eq:CN08_te}\\
& \tau_m^{-1} \simeq \frac{2.7 + 1.1 \, p}{2} h^{2} 
\frac{1 - \left( \frac{\hat{e}}{2.02} \right)^4}
{1 + \left( \frac{\hat{e}}{2.25} \right)^{1/2} + \left( \frac{\hat{e}}{2.84} \right)^6}
\; t_{\rm wave}^{-1}, \label{eq:CN08_tm}
\end{align}
where they added 
the factor $(2.7 + 1.1 \, p)$ corresponding to Eq.~(\ref{eq:facC}) with $q=0$
to $\tau_m^{-1}$, 
although the disc model they used for the fitting had $p=1/2$ and $q=1$.
Figure~\ref{fig:mig_time} shows the results of 
Eqs.~(\ref{eq:CN08_te}) and (\ref{eq:CN08_tm}) with $p=1/2$.
In the supersonic regime
in \citet{Cresswell08}'s result, in contrast to \citet{Papaloizou00}, $\tau_a^{-1}$ is always positive, which is in agreement with hydrodynamical simulations \citep{Cresswell07, Bitsch10},
while $\tau_e^{-1}$ is similar
for both \citet{Cresswell08} and \citet{Papaloizou00}.
Because Figure~\ref{fig:mig_time} shows that $\tau_e \ll \tau_a$ 
as long as $e^2 \ll 1$, $e$ is generally reduced before 
a planetary embryo significantly migrates.
However, when $e$ is continuously excited by planet-planet resonant perturbations or secular perturbations by a giant planet, the different values of $\tau_a$ for $e \ga h$ should make a difference in dynamics of planetary 
embryos and the formation of planets. 

While \citet{Cresswell08} used the force formula by \citet{Papaloizou00} (Eq.~(\ref{eq:dvdtPL00})),
\citet{Coleman14} used the following formulae:
\begin{equation}
\frac{d\vct{v}}{dt} = - \frac{\vct{v}}{\tau_m} -  \frac{v_r}{\tau_e} \vct{e}_r - \frac{0.5(v_\theta - v_{K,a}) }{\tau_e} \vct{e}_\theta,
\label{eq:dvdtCN14}
\end{equation}
where $v_{K,a}$ is Keplerian velocity at the semimajor axis $a$.
They still adopted the finite $e$ corrections listed in Eqs.~(\ref{eq:CN08_te}) and (\ref{eq:CN08_tm}).

\citet{Kominami05}, \citet{Kominami06}, and \citet{Ogihara07} adopted a different form of the equations of motion as
\begin{equation}
\frac{d\vct{v}}{dt} = - \frac{v_K}{\tau_{m,e=0}} \vct{e}_\theta
-  \frac{ f_{r,\rm TW} }{0.780 \, \tau_e} \vct{e}_r 
- \frac{ f_{\theta, \rm TW}}{0.780 \, \tau_e} \vct{e}_\theta,
\label{eq:dvdtKO}
\end{equation}
where $f_{r,\rm TW}$ and $f_{\theta, \rm TW}$ are 
given by Eqs.~(\ref{eq:TWr}) and (\ref{eq:TWtheta}), and $v_K$ is
defined at instantaneous radius $r$. 
Because \citet{Tanaka04} did not give a force formula for $a$ damping,
they added the first term as a force for orbital migration.

We will discuss the different forms of equations of motion, Eqs.~(\ref{eq:dvdtPL00}), (\ref{eq:dvdtCN14}), and 
Eq.~(\ref{eq:dvdtKO}), in section 4.

\section{Migration and Eccentricity Damping Prescriptions Based on Dynamical Friction}


\subsection{Derivation of simple formulas based on dynamical friction}

 Hydrodynamical simulations \citep[e.g.][]{Bitsch10} show that, 
 in the subsonic case, planet-disc interactions occur mostly 
 through spiral density waves, whereas in the supersonic case it is through dynamical friction. 
In the supersonic case, the relative velocity for dynamical friction
is mainly caused by the eccentricity of the planetary orbit.  However, even in the subsonic case, dynamical friction contributes to 
planet-disc interactions in addition to the torques from the density waves.

\citet{Muto11} analytically derived the specific force
of 2D dynamical friction  
from uniform gas flow to a planetary embryo, which is given by
\begin{equation}
\vct{F}_{\rm DF,MTI} \simeq
\left\{ \begin{array}{ll}
\displaystyle 
 - \pi \, \frac{\Delta \vct{v}}{t_{{\rm wave},r}} \tilde{\epsilon}^{-2} \alpha
& [{\rm for}\; \Delta v \ll c_s] 
 \\
\displaystyle 
 - 2 \pi \, \frac{\Delta \vct{v}}{t_{{\rm wave},r}}\left( \frac{\Delta v}{c_s} \right)^{-3}
\tilde{\epsilon}^{-1}
 & [{\rm for}\; \Delta v \gg c_s], 
\end{array}
\right.
\label{eq:FDF_Muto}
\end{equation}
where $ \tilde{\epsilon}= \epsilon/0.5H$,
$t_{{\rm wave},r}$ is defined by Eq.~(\ref{eq:twave}) at instantaneous $r$,
and the effective ``viscosity" parameter $\alpha$ is discussed below.
Recently, \citet{Sanchez19} showed through hydrodynamical simulations
that \citet{Muto11}'s formula in the supersonic case is valid for $0.1 < e < 0.6$
(also see \citet{Vincente19}).

\citet{Muto11}'s subsonic formula needs careful treatment, 
because it depends on the uncertain parameter $\alpha$.
\citet{Rephaeli80} showed that the dynamical friction from inviscid gas flow is zero,
because the flow on frontside of the body and that on the backside are symmetric in steady state.
\citet{Muto11} introduced the viscosity of the gas flow,
resulting in asymmetry and non-zero dynamical friction.
Another effect to produce the asymmetry is non-steadiness of the gas flow.
Because the timescale for the steady state of the inviscid gas flow to be established is infinite, \citet{Ostriker99} found that the gas flow established on finite timescale
is asymmetric and the net dynamical friction exists.
In the planet-disc interactions, the gas flow around the planetary embryo is not steady
because of the epicyclic motion of the planetary embryo. 
In this paper, we determine the value of $\alpha$ 
in the subsonic formula in Eq.~(\ref{eq:FDF_Muto})
by \citet{Tanaka04}'s formula; in other words, for the subsonic formula, 
we use the part of \citet{Tanaka04}'s force formula that
contributes to secular decrease of $e$, as explained below.
Here, we infer that 
$\alpha$ represents the degree of non-steadiness
of the gas flow interacting with the planetary embryo
due to the epicyclic motion of the planetary embryo, 
rather than the strength of turbulence of disc gas flow,
although the derived subsonic formula does not depend on
the physical interpretation of $\alpha$.

\citet{Tanaka04} derived 
the $e$-damping rate and force for a planetary embryo in epicycle motion
with $e < h$ in the disc flow with the shearing-sheet approximation, through 3D linear calculation
(Eqs.~(\ref{eq:TWr}) and (\ref{eq:TWtheta})).
Since the numerical factor obtained by \citet{Tanaka04} is rigorous,
we will adopt the equivalent force formula to \citet{Tanaka04}'s one in the subsonic case.
As we will show below,
the force formula is equivalent to \citet{Muto11}'s formula in the subsonic case, with $\alpha \simeq (0.78/\pi)\tilde{\epsilon}^{2}$,
which represents the non-steady gas flow due to the epicycle motion but not 
a real turbulent viscosity.
Interestingly, with the estimated value of $\alpha$,  
\citet{Muto11}'s formula is consistent with
Chandrasekhar's dynamical friction formula for stellar dynamics
both in supersonic and subsonic cases
(Appendix B), despite the interactions with disc gas
are different from those with particles (field stars) because of gas pressure in the subsonic case.
We will also show below that the contribution to the $e$-damping
due to density waves that are linked to gas pressure
vanishes on an orbit average.  

\citet{Tanaka04}'s formulas (Eqs.~(\ref{eq:TWr}) and (\ref{eq:TWtheta})) are rewritten as
\begin{align}
& f_{r,\rm TW} \simeq 
\left\{ - 0.780 v_r + \left[0.114 v_y + 0.956 v_r \right] \right\}, 
\label{eq:frTW} \\
& f_{\theta,\rm TW} \simeq 
\left\{ - 0.780 v_y + \left[- 0.956 v_y + 0.325 v_r \right] \right\}, \label{eq:fyTW}
\end{align}
where $v_y = (v_\theta - r\Omega)$, and we split their formula into the dynamical friction parts (the first terms in the r.h.s.) that are anti-parallel to $\Delta \vct{v}$, 
and the residual parts (the second and third terms).
The dynamical friction force is always anti-parallel to $\Delta \vct{v}$ in \citet{Muto11}'s formula (Eq.~(\ref{eq:FDF_Muto})).
In the subsonic case, density waves are excited.
The non-parallel components would correspond to the forces from density waves, 
and they vanish on the orbit average as shown below.

In the local limit (Hill approximation), 
$v_r = e v_K \sin (\Omega t)$ and $v_y=(e/2) v_K \cos (\Omega t)$
\citep[e.g.][]{Henon86}, respectively,
so that the ``energy" of velocity dispersion is given by
\begin{equation}
E_e = \frac{1}{2} e^2 v_K^2 = \frac{1}{2}(v_r^2 + (2v_y)^2).
\label{eq:Ee}
\end{equation}
The power $dE_e/dt$ is
\begin{align}
\frac{dE_e}{dt} = & \; v_r  \frac{dv_r}{dt} +  4 v_y  \frac{dv_y}{dt}
= t_{\rm wave}^{-1} (v_r \, f_{r,\rm TW} + 4 v_y \, f_{\theta,\rm TW}) 
 \nonumber\\
\simeq & \; t_{\rm wave}^{-1} 
 \left[- 0.780 \,v_r^2 + 0.956 \,v_r^2 - 0.780 \times 4\, v_y^2  - 0.956 \times 4 \,v_y^2 \right] \nonumber\\
  & + t_{\rm wave}^{-1} \left[0.114 \,v_y v_r + 0.325 \times 4\,v_r v_y \right].  
\label{eq:DF1}
\end{align} 
Performing an orbit average, the cross terms with $v_y v_r$ vanish and we obtain
\begin{align}
\langle \frac{d E_e}{dt} \rangle
 \sim - t_{\rm wave}^{-1} 
\left\{ 0.780 \times 2\, E_e - 0.956[\langle v_r^2 \rangle  - \langle 4 v_y^2\rangle] \right\}. \label{eq:DF2}
\end{align} 
Because the orbit averages of $v_r = e v_K \sin (\Omega t)$ and $v_y=(e/2) v_K \cos (\Omega t)$ satisfy
\begin{equation}
\langle v_r^2 \rangle \simeq \langle 4 v_y^2 \rangle,
\label{eq:vr4vy}
\end{equation}
the term in the square brackets in Eq.~(\ref{eq:DF2}) vanishes.

Thus, the eccentricity is damped by the density wakes associated with dynamical friction, which are the first terms in the right hand side of Eqs.~(\ref{eq:frTW}) and (\ref{eq:fyTW}).
Note that 
\begin{align}
\frac{1}{e}\frac{de}{dt} = \frac{1}{2\,E_e}\frac{dE_e}{dt} \simeq -0.780 \, t_{\rm wave}^{-1},
\end{align}
which perfectly agrees with the $e$-damping rate that \citet{Tanaka04} derived in a different way.
The non-parallel parts just raise oscillation of $e$ and do not produce net change in $e$.
We adopt the secular change parts in \citet{Tanaka04}'s force formulae as
the $e$-damping force in subsonic limit,
\begin{align}
\vct{F}_{\rm DF,TW} = - 0.780\, \frac{\vct\Delta {v}}{t_{\rm wave}}. 
\label{eq:DF4}
\end{align} 
This force formula is equivalent to
\citet{Muto11}'s subsonic dynamical friction formula (Eq.~(\ref{eq:FDF_Muto}))
with $\alpha \simeq (0.78/\pi)\tilde{\epsilon}^{2}$.

The relative velocity of close encounters between a planetary embryo and gas that is defined by
\begin{equation}
\Delta \vct{v} = \vct{v} - \vct{v}_{\rm gas} 
= v_r \vct{e}_r + \left[v_\theta - r (1-\eta) \Omega_K \right] \vct{e}_\theta,
\label{eq:Del_v0}
\end{equation}
where we included the factor of $\eta \equiv - (h^2/2)(d\ln P/d\ln r)$ 
that represents the deviation of disc gas velocity from Keplerian velocity; 
for discs with smooth surface density radial distribution, $\eta \sim 1.3 \, h^2 \ll 1$. 
When $e > \eta$, 
the deviation from local circular Keplerian velocity
is given by $\Delta v \sim e v_K = e r \Omega_K$.
In this case, the $e$-damping rate is
\begin{align}
\tau_{e, \rm sub}^{-1} & \simeq - \frac{F_{\rm DF,TW}}{\Delta v} \simeq 0.780 
\, t_{\rm wave}^{-1}.
\label{eq:tau_e_DF_sub}
\end{align}

In the supersonic regime, the uncertain parameter $\alpha$ is not included
in Eq.~(\ref{eq:FDF_Muto}).
We use Eq.~(\ref{eq:FDF_Muto}) for supersonic regime,
\begin{equation}
\vct{F}_{\rm DF,MTI,sup} \simeq
\displaystyle 
 - 2 \pi \, \frac{\Delta \vct{v}}{t_{{\rm wave},r}}\left( \frac{\Delta v}{c_s} \right)^{-3}
\tilde{\epsilon}^{-1}.
\label{eq:FDF_sup}
\end{equation}
We used the asymptotic function for $e/h \gg 1$ in the supersonic case, but kept $e^2 \ll 1$.

In this relatively high $e$ regime,
the relation between the eccentricity damping rate $\tau_e^{-1}$ and 
$\vct{F}_{\rm DF,MTI,sup}$ is not simple, because the change in $r$ 
within one orbit is not negligible.
\citet{Muto11} integrated $\vct{F}_{\rm DF,MTI,sup}$ in one orbit
to obtain $\tau_e^{-1}$.
They showed results for 
$M_p/M_*=10^{-6}$, $\Sigma r^2/M_*=10^{-4}$, 
$\tilde{\epsilon} = 1$, and $h=0.05$ at 1 au 
($t_{\rm wave} \simeq 10^4$ yr) in their Figure 9,
which are fitted for $e \la 0.5$ as
\begin{equation}
\tau_{e, \rm sup}^{-1} \simeq 12 \; \hat{e}^{-3} \, t_{\rm wave}^{-1},
\label{eq:tau_e_DF_super}
\end{equation}  
where $t_{\rm wave}^{-1}$ is defined at the semimajor axis.
In this regime, $\tau_e^{-1}$ depends on the disc parameters only through 
local values, $t_{\rm wave}$, but is almost independent of their radial gradients, $p$ and $q$. 
The numerical factor is increased 
by a factor of 2 by the variation of $r$, compared to
Eq.~(\ref{eq:FDF_Muto}) in the supersonic case. 

Now we combine $\vct{F}_{\rm DF,MTI,sup}$ in the supersonic limit (Eq.~\ref{eq:FDF_Muto}) and 
$\vct{F}_{\rm DF,TW}$ in the subsonic limit (Eq.~\ref{eq:DF4}).
Although there are many ways to combine the two limits,
we here adopt a simple summation of the timescales.
We will discuss on this issue again in section 3.2.

The timescale summation of the two limits
of the damping rates and forces
(Eqs.~(\ref{eq:tau_e_DF_sub}), (\ref{eq:tau_e_DF_super}), (\ref{eq:DF4}), and (\ref{eq:FDF_sup})), 
we obtain
\begin{align}
\tau_e^{-1} & \simeq 0.780 \left(1 + \frac{1}{15} \hat{e}^{3} \right)^{-1} \, t_{\rm wave}^{-1},
\label{eq:tau_e_DF}
\end{align}
and
\begin{equation}
\vct{F}_{\rm DF} \simeq - 0.780 \, \frac{\Delta \vct{v}}{t_{{\rm wave,}  r}}
\left[1 + \frac{1}{8} \left( \frac{\Delta v}{c_s} \right)^{3} \right]^{-1},
\label{eq:DFmuto}
\end{equation}
where we adopted $\tilde{\epsilon}=1$ in Eq.~(\ref{eq:FDF_sup})
and $t_{{\rm wave,} r}$ is defined at instantaneous radius $r$.
In the limits of $\tilde{e} \ll 1$ and $\tilde{e} \gg 1$,
$\tau_e^{-1}$ is reduced to Eq.~(\ref{eq:tau_e_DF_sub}) and
Eq.~(\ref{eq:tau_e_DF_super}), respectively.
In the limits of $\Delta v \ll c_s$ and $\Delta v \gg c_s$,
$\vct{F}_{\rm DF}$ is reduced to Eq.~(\ref{eq:DF4}) and
 Eq.~(\ref{eq:FDF_sup}), respectively. 
The difference in the numerical factor for the high $e$-limit (supersonic limit)
in $\tau_e^{-1}$ from that in $\vct{F}_{\rm DF}$
comes from the variation in $r$ in during one orbit.

\citet{Muto11} showed that $\tau_a^{-1}$ determined by the dynamical friction formula
is positive in the supersonic regime. 
To proceed, we also use
the same connection of the supersonic and subsonic limit expressions of $\tau_a^{-1}$
(Because $\tau_m$ changes a sign at $e\sim h$, the same connection of the two limits for this quantity
does not make sense).
\citet{Muto11} showed that $\tau_a^{-1}$ depends on the gradients of the disc parameters, $p$ and $q$, 
while the $p$, $q$-dependences of $\tau_e^{-1}$ are negligible. 

In the supersonic limit, their results of a numerically integrated analytical equation with 
$M_p/M_*=10^{-6}$, $\Sigma r^2/M_*=10^{-4}$, $\epsilon = 0.5H$, and $h=0.05$ at 1 au (their Figure 9) are fitted as
\begin{equation}
\tau_a^{-1} = C_{\rm M} \hat{e}^{-1} \, h^2 \, t_{\rm wave}^{-1},
\label{eq:adot_Tanaka}
\end{equation}  
where $C_{\rm M} = 6 (2p-q+2)$.
The values of $C_{\rm M}$ in different disc models 
are listed in Table 1.

In the subsonic limit, we here adopt the isothermal migration rate by \citet{Tanaka02}, 
\begin{align}
\tau_a^{-1} & = C_{\rm T} h^{2} \; t_{\rm wave}^{-1}.
\label{eq:tau_a3}
\end{align}
The formulae with the non-isothermal migration by \citet{Paardekooper11}
is discussed in Appendix C.
As we will show in section 4,
the dynamical friction argument can evaluate 
$\tau_a^{-1}$ in the subsonic case, but only by order of magnitude.
Furthermore, in the subsonic case, the contribution by density waves 
to $\tau_a^{-1}$ does not cancel through orbit average and it dominates in the non-isothermal case
(Appendix C), while the contribution to $\tau_e^{-1}$ cancels as we showed before.
As such, we adopt Eq.~(\ref{eq:adot_Tanaka}) for the subsonic limit.

We combine these two limits to obtain
\begin{align}
\tau_a^{-1} & \simeq C_{\rm T} h^{2} 
\left( 1 + \frac{C_{\rm T}}{C_{\rm M}} \hat{e} \right)^{-1}t_{\rm wave}^{-1}.
\label{eq:tau_a_DF}
\end{align}
From Eqs.~(\ref{eq:tau_e_DF})  and (\ref{eq:tau_a_DF}),
\begin{align}
\tau_m^{-1} & = \frac{1}{2}\tau_a^{-1} - \frac{e^2}{1-e^2}\tau_e^{-1} \simeq
\frac{1}{2}\tau_a^{-1} - e^2 \tau_e^{-1} \nonumber \\
 & \simeq h^{2}  
 \left( \frac{C_{\rm T}}{2} \frac{1}{1 + \frac{C_{\rm T}}{C_{\rm M}} 
 \hat{e}} 
 - 0.78\,  \hat{e}^2 \frac{1}{1 + \frac{1}{15} \hat{e}^3} \right)
 t_{\rm wave}^{-1}.
\label{eq:tau_m_DF}
\end{align}
In the subsonic limit the first term in the bracket dominates and $\tau_m^{-1} > 0$.
In the supersonic limit we have
\begin{align}
\tau_m^{-1} & \simeq h^{2} \, \hat{e}^{-1}
\left( \frac{C_{\rm M}}{2} - 12 \right)\, t_{\rm wave}^{-1} \nonumber\\
  & \simeq - 3(2 - 2\,p + q) \, h^{2} \hat{e}^{-1}
 t_{\rm wave}^{-1}.
\label{eq:tau_m_DF_super}
\end{align}
In the steady accretion discs,
$3\pi \Sigma \nu \propto r^{-p - q +3/2}$ is a constant of $r$,
so that $p+q=3/2$.
In this case, the numerical factor in the above equation,
$- 3(2 - 2\,p + q) = - 3(7/2 - 3\,p)$, is negative both for 
irradiation dominated ($p=15/14$) and viscous-heating dominated ($p=3/5$) regimes
\citep{Ida16}.
Therefore, in the supersonic case, $\tau_m^{-1} < 0$ 
(causing an increase in the angular momentum), while $e$ and $a$ are always damped ($\tau_e^{-1}, \tau_a^{-1} > 0$), as long as the locally isothermal case is considered.

\subsection{Comparison with \citet{Papaloizou00} and \citet{Cresswell08}}

\begin{table}
\begin{tabular}{cccc}\hline
Numerical factors & Nominal & PL00 & CN08 \\ \hline \hline 
$C_{\rm M} =$ & & & \\
$6 (2\,p - \,q + 2)$ & 21 & 24 & 12 \\ \hline
$C_{\rm T}=$ & & & \\
$2.73+1.08 \, p + 0.87 \,q$ & 4.25 & 5.22 & 4.14 \\ \hline
$C_{\rm P}=$ & & & \\
$2.50 - 0.1 \,p + 1.7\, q $ &  3.25 & 4.05 & 4.15\\ \hline
\end{tabular} 
\caption{Numerical coefficients.  Here $p = - d \ln \Sigma/ d\ln r$ and $q = - d \ln T/ d\ln r$. Numerical values are given for 
the nominal steady disc ($p=1, \,q=0.5$), which is
close to an irradiation dominated disc \citep{Ida16},
PL00's disc with $(p,q)=(1.5,1)$, and CN08's disc with $(p,q)=(0.5,1)$, respectively.
Note that the PL00's and CN08's discs are not steady accretion ones.}
\end{table}

In Figure~\ref{fig:mig_time}, our new formulae given above based on dynamical friction are compared with
the formulae by \citet{Papaloizou00} and \citet{Cresswell08}
for $h = 0.05$ and $\epsilon = 0.5\,H$.
\citet{Papaloizou00} and \citet{Cresswell08} used
fixed discs $(p, q)=(1.5, 1.0)$ and $(0.5, 1.0)$,
while we plot our results for the nominal irradiative steady accretion disc
\footnote{Although \citet{Ida16} showed for the irradiative regime 
$(p, q)=(15/14, 3/7)$, we used a more simple values, $(p, q)=(1.0, 0.5)$.}
with $(p, q)=(1.0, 0.5)$ and CN08's disc with $(0.5, 1.0)$.
As shown in Table 1, the differences in $C_{\rm M}$ and $C_{\rm T}$ (and also $C_{\rm P}$
in the next subsection) for the different $(p,q)$ are within a factor of 2.

Our formulation with both the nominal disc
and CN08's disc is similar to that of \citet{Cresswell08} that was obtained by
a fitting of hydrodynamical simulations.
The migration is always inward ($\tau_a^{-1} > 0$).
The derivation of our formulae is much simpler and more intuitive than the 
gravitational potential expansion done by 
\citet{Papaloizou00} and the fitting of hydrodynamical simulation results done by \citet{Cresswell08}.
Our formulae also for the first time explicitly show the dependence on the disc parameters through 
$C_{\rm T} \,(= 2.73+1.08 \, p + 0.87 \,q)$ and $C_{\rm M} \,(= 6 (2p -q + 2))$. 

The small peaks at $e/h \sim$ a few in $\tau_e^{-1}$ and $\tau_a^{-1}$ in  the \citet{Cresswell08}'s formulae
reflect the results of hydrodynamical simulations \citep{Cresswell08, Bitsch10}.
We note that the dynamical friction formulae by \citet{Muto11}
include a divergence at $\Delta v = c_s$, which corresponds to $e \sim h$. 
Because the divergence is so sharp that it should be
smoothed out by non-linear effects and orbit averaging where $\Delta v$ changes a factor of a few in one orbit,
$\tau_e^{-1}$ and $\tau_a^{-1}$ should have some peak at $e \sim h$.  If we take this effect into account,
our formulae become more similar to those of \citet{Cresswell08}. 
However, because it may make our simple, intuitive formulae more complicated and requires additional manipulation,
we decided not to do so in the present paper.

\section{Equations of Motion}

So far, we have mainly discussed 
the rates of change (the inverse of  the timescales) in $e$, $a$, and $\ell$.
In addition to the timescales, 
how to implement them in the equations of motion for N-body simulations 
have been proposed in different ways.

The proposed implementation to the equations of motion 
by \citet{Papaloizou00} (Eq.~\ref{eq:dvdtPL00}), \citet{Coleman14} (Eq.~\ref{eq:dvdtCN14}),
and \citet{Kominami06} (Eq.~\ref{eq:dvdtKO}) are given respectively by
\begin{align}
\frac{d\vct{v}}{dt} & = - \frac{\vct{v}}{\tau_m} -  2\frac{v_r}{\tau_e} \vct{e}_r,
\label{eq:dvdtPL00b} \\
\frac{d\vct{v}}{dt} & = - \frac{\vct{v}}{\tau_m} -  \frac{v_r}{\tau_e} \vct{e}_r - \frac{0.5(v_\theta - v_{K,a}) }{\tau_e} \vct{e}_\theta, 
\label{eq:dvdtCN14b} \\
\frac{d\vct{v}}{dt} & = - \frac{v_K}{\tau_{m,e=0}} \vct{e}_\theta
+ \frac{f_{r,\rm TW} }{0.78 \,\tau_e} \vct{e}_r 
+ \frac{f_{\theta, \rm TW}}{0.78\,\tau_e} \vct{e}_\theta,
\label{eq:dvdtKOb}
\end{align}
where $v_{K,a}$ is Keplerian velocity at the planetary semimajor axis $a$,
while $v_{K}$ represents Keplerian velocity at instantaneous radius $r$ in this paper.
Except for \citet{Tanaka04}'s eccentricity damping force, 
all of these equations of motion were given a priori by using the timescales without their derivation. 
Here we discuss the consistency of these force formulae.

In the dynamical friction formulation the form of the equations of motion is straightforward:
\begin{align}
\frac{d\vct{v}}{dt} & = - \frac{\Delta \vct{v}}{\tau_e} = - \frac{v_r}{\tau_e} \vct{e}_r - \frac{v_\theta - r (1-\eta) \Omega_K}{\tau_e} \vct{e}_\theta. 
\label{eq:dvdtF2}
\end{align}
We can split the second term proportional to $\vct{e}_\theta$ into 
the migration and the eccentricity damping parts as
\begin{align}
\frac{d\vct{v}}{dt} & = -  \frac{v_K}{\tau_e \,\eta^{-1}}\vct{e}_\theta
- \frac{v_r}{\tau_e} \vct{e}_r - \frac{v_\theta - v_K}{\tau_e} \vct{e}_\theta.
\label{eq:dvdtF3}
\end{align}

In this paper, we consider discs with smooth surface density distribution
(The discussion here cannot be applied for the discs with gaps and rings).
In this case, $\eta \sim h^2$ and $\tau_e \,\eta^{-1} \sim \tau_{m,e=0}$ in the subsonic case.
As we discussed in section 3.1, only the parts related to dynamical friction
actually contribute to the eccentricity damping.
Therefore, the equations of motion by 
\citet{Kominami06} (Eq.~(\ref{eq:dvdtKOb})) are justified for the subsonic case.
However, Eq.~(\ref{eq:dvdtKOb}) cannot be applied for the supersonic regime. 
Furthermore, as we pointed out in section 3.2,
the contribution by density waves to migration 
is also important in the subsonic case, in particular,
in the non-isothermal discs (Appendix C).

Therefore, as the equations of motion that can be applied in both 
subsonic and supersonic regimes, we propose
\begin{align}
\frac{d\vct{v}}{dt} & = -  \frac{v_K}{2 \tau_a}\vct{e}_\theta 
- \frac{v_r}{\tau_e} \vct{e}_r - \frac{v_\theta - v_K}{\tau_e} \vct{e}_\theta.
\label{eq:dvdtF4}
\end{align}
If inclination is considered, $- (v_z/\tau_i) \vct{e}_z$ is added,
where  $\tau_i$ is given in Appendix D.

Next we consider the form for equations of motion shown in Eqs.~(\ref{eq:dvdtPL00b}) and (\ref{eq:dvdtCN14b}). 
We assume the following form,
\begin{equation}
\frac{d\vct{v}}{dt} = - \frac{\vct{v}}{\tau} -  A_r \frac{v_r}{\tau_e} \vct{e}_r
-  A_\theta \frac{v_\theta - v_K}{\tau_e} \vct{e}_\theta,
\label{eq:dvdt}
\end{equation}
where $\tau$ is an unknown timescale and
$A_r$ and $A_\theta$ are unknown factors of $\sim O(1)$.
In the below, we derive $\tau$, $A_r$, and $A_\theta$.
By the definition of $\tau_m$,
\begin{align}
- \frac{\vct{r} \times \vct{v}}{\tau_m} & = \frac{d(\vct{r} \times \vct{v})}{dt} =
\vct{r} \times \frac{d\vct{v}}{dt}.
\label{eq:rv_tau_m}
\end{align}
Taking the cross product of both sides of Eq.~(\ref{eq:dvdt}), we have
\begin{align}
\vct{r} \times \frac{d\vct{v}}{dt} & = - \frac{\vct{r} \times \vct{v}}{\tau} 
-  A_\theta \frac{v_\theta - v_K}{\tau_e} \vct{r} \times \vct{e}_\theta \nonumber \\
 & = - \frac{\vct{r} \times \vct{v}}{\tau} 
-  A_\theta \frac{v_\theta - v_K}{v_\theta \tau_e} \vct{r} \times \vct{v},
\label{eq:dvdtPL00c}
\end{align}
because $\vct{v} = v_r \vct{e}_r + v_\theta \vct{e}_\theta$.
From Eqs.~(\ref{eq:rv_tau_m}) and (\ref{eq:dvdtPL00c}) we obtain
\begin{align}
- \frac{1}{\tau_m} 
= - \frac{1}{\tau} -  A_\theta \frac{(v_\theta - v_K)}{v_\theta}\frac{1}{\tau_e}.
\end{align}
Substituting this into Eq.~(\ref{eq:dvdt}), we arrive at
\begin{align}
\frac{d\vct{v}}{dt} & = - \frac{\vct{v}}{\tau_m} 
+ A_\theta \frac{(v_\theta - v_K)}{v_\theta} \frac{\vct{v}}{\tau_e}
-  A_\theta \frac{v_\theta - v_K}{\tau_e} \vct{e}_\theta -  A_r \frac{v_r}{\tau_e}\vct{e}_r 
\nonumber \\
 &  = - \frac{\vct{v}}{\tau_m} 
-  \left( -A_\theta \frac{v_\theta - v_K}{v_\theta}+A_r\right) \frac{v_r}{\tau_e} \vct{e}_r.
\label{eq:dvdt3}
\end{align}
Taking the dot product of both sides of this equation with $\vct{v}$, we obtain
the change rate of the orbital energy ($E=\vct{v}^2/2 - GM_*/r$),
\begin{align}
 & \frac{dE}{dt} = \frac{d (\vct{v}^2/2)}{dt} = 
 - \frac{\vct{v}^2}{\tau_m} - \left( -A_\theta \frac{v_\theta - v_K}{v_\theta}+A_r\right) \frac{v_r^2}{\tau_e}.
\label{eq:dvdt4}
\end{align}
Because $E$ is given by the semimajor axis $a$ as $E = - GM_*/2a$,
we have $dE/dt = - v_{K}^2/(2 \tau_a)$, so that Eq.~(\ref{eq:dvdt4}) becomes
\begin{align}
\frac{v_{K}^2}{2\tau_a} & = \frac{v^2}{\tau_m}
+ \left( -A_\theta \frac{v_\theta - v_K}{v_\theta}+A_r\right) \frac{v_r^2}{\tau_e}.
\label{eq:dvdt5}
\end{align}
From Eq.~(\ref{eq:ldot2}), we can write in the case of $e^2 \ll 1$,
\begin{align}
\frac{v_{K}^2}{2\tau_a} & \simeq \frac{v_{K}^2}{\tau_m}
+ \frac{e^2 v_{K}^2}{\tau_e} \simeq \frac{v^2}{\tau_m}
+ \frac{2 v_r^2}{\tau_e},
\label{eq:dvdt6}
\end{align}
where  we used Eqs.~(\ref{eq:Ee}) and (\ref{eq:vr4vy})  in the last equation.
Comparing Eq.~(\ref{eq:dvdt5}) and Eq.~(\ref{eq:dvdt6}),
$\left( -A_\theta (v_\theta - v_{K})/v_\theta)+A_r\right) \simeq 2$
and Eq.~(\ref{eq:dvdt3}) is approximated as
\begin{align}
\frac{d\vct{v}}{dt} &  = - \frac{\vct{v}}{\tau_m} 
-  2 \frac{v_r}{\tau_e} \vct{e}_r,
\label{eq:dvdt7}
\end{align}
which is identical to the formulation of \citet{Papaloizou00}'s 
equations of motion, Eq.~(\ref{eq:dvdtPL00b}).
Note that the azimuthal component of eccentricity damping
is included in the fist term, $\vct{v}/\tau_m$, which is not visible
in Eq.~(\ref{eq:dvdt7}). 
In the supersonic case, the 1st term of Eq.~(\ref{eq:dvdt7}) changes the sign.
However, due to the effect of the 2nd term, the orbital migration is still inward as we discussed.

It is not clear if the form of Eq.~(\ref{eq:dvdtCN14b}),
in which the azimuthal component of eccentricity damping is explicitly applied, is justified.
Note that \citet{Coleman16a} and \citet{Coleman16b} adopted Eq.~(\ref{eq:dvdt7}) rather than Eq.~(\ref{eq:dvdtCN14b}) used in \citet{Coleman14}. 

In summary, in our opinion the safest equation of motion to use is Eq.~(\ref{eq:dvdtF4}),
because it is most consistent 
with the simple argument based on dynamical friction. 
We can reproduce the equations of motion (Eq.~(\ref{eq:dvdt7})) adopted by \citet{Papaloizou00} 
under the condition of $e^2 \ll 1$ as in Eq.~(\ref{eq:dvdt6}), but we are unable to reproduce Eq.~(\ref{eq:dvdtCN14b}), the equations of motion adopted by \citet{Coleman14}.

\section{Conclusions}

Disc-planet interactions often occur in supersonic regime
in the planet accretion processes in the protoplanetary disc,
as discussed in section 2.1.
Although the planet-disc interaction in the supersonic regime is important, 
the prescriptions proposed so far \citep{Papaloizou00, Tanaka04,
Kominami06, Cresswell07, Coleman14} 
seem inconsistent with one another and are sometimes confusing.
 
In this paper, after comparing the existing prescriptions in detail,
we have proposed a simple prescription for the planet-disc interactions 
that is applicable for both supersonic and subsonic cases.
Because our derivation is based on dynamical friction formulae,
the derivation is intuitively understood.

Our prescription is summarized as follows.
The safest equations of motion are (Eqs.~(\ref{eq:dvdtF4})):
\begin{align}
\frac{d\vct{v}}{dt} & = -  \frac{v_{K}}{2 \tau_a}\vct{e}_\theta 
- \frac{v_r}{\tau_e} \vct{e}_r - \frac{v_\theta - v_{K}}{\tau_e} \vct{e}_\theta.
\label{eq:dvdtF6}
\end{align}
The damping timescales of the orbital eccentricity, semimajor axis, and angular momentum
are given by  (Eqs.~(\ref{eq:tau_e_DF}), (\ref{eq:tau_a_DF}), and (\ref{eq:tau_m}))
\begin{align}
\tau_e^{-1} & \simeq 0.780\, \left(1 + \frac{1}{15} \left(\frac{e}{h}\right)^{3} \right)^{-1} t_{\rm wave}^{-1} , 
\label{eq:taue_f2}\\
\tau_a^{-1} & \simeq C_{\rm T} h^{2}
\left( 1 + \frac{C_{\rm T}}{C_{\rm M}} \frac{e}{h} \right)^{-1}t_{\rm wave}^{-1},\label{eq:taua_f2}\\
\tau_m^{-1} & \simeq \frac{1}{2}\tau_a^{-1} - e^2 \tau_e^{-1},\label{eq:taum_f2}
\end{align}
where 
$C_{\rm T} = 2.73+1.08\,p+0.87\,q$, $C_{\rm M} = 6 (2\,p - \,q + 2)$, 
$p = - d\ln \Sigma/d\ln r$, $q = - d\ln T/d\ln r$,  
\begin{equation}
t_{\rm wave}^{-1} = \left(\frac{M_p}{M_*}\right)\left(\frac{\Sigma r^2}{M_*}\right)
h^{-4}\Omega_K,
\end{equation}
and $t_{\rm wave}^{-1}$ is evaluated at 
the semimajor axis $a$.  
In the non-isothermal case, Eqs.~(\ref{eq:non_iso_e}), (\ref{eq:non_iso_a}), and
(\ref{eq:non_iso_m}) in Appendix C should be used,
instead of Eqs.~(\ref{eq:taue_f2}), and (\ref{eq:taua_f2}), and (\ref{eq:taum_f2}).
The formulae with non-zero inclination $i$ ($< h$) are given in Appendix D.

The formula by \citet{Cresswell07} is obtained by a fitting with hydrodynamical simulations
and takes detailed features into account, such as an 
enhancement of the interactions near $e \sim 2{\rm -}3 \,h$.
However, because only one disc model was used for the hydrodynamical simulations,
it is not clear how the detailed features depend on the disc parameters.
Although our formulae do not reproduce the enhancement 
near the subsonic-supersonic boundary,
they have explicit dependences on the disc parameters,  $p$ and $q$.
Note, however, that 
we need to be careful when we apply this to the disc structure with local variations that have radial scales less than the disc scale height ($H$), because all of \citet{Tanaka02}, \citet{Tanaka04}, and \citet{Muto11}, which we used to derive the dependences 
on $p$ and $q$, assumed that local uniformity on the scale of $H$.

N-body simulation is one of the most powerful tools to study formation of 
close-in multiple super-Earths that are observed commonly in exoplanetary discs.
As \citet{Brasser18} demonstrated,
the resonant trapping and collisions between the planetary embryos near the disc inner edge sensitively depend on the adopted prescription of the planet-disc interactions,
because $e$ of the planetary embryos in the resonant chains are excited to $\sim h$,
which is near the subsonic-supersonic boundary. 
\footnote{Note that at $e \sim$ a few $h$, 
all the prescriptions show that the planet-disc interactions
increases the angular momentum ($\sqrt{GM_* a (1-e^2)}$).
On the other hand, our and \citet{Cresswell08}'s prescriptions
show the inward migration (decrease in $a$).
For the inward migration to actually occur, $e$ must be reduced by the planet-disc interactions and it must be continuously excited by the same mechanism such as the resonant perturbations between the planetary embryos or secular perturbations from a giant planet.
}
Our new prescription will reduce the uncertainty due to the different prescriptions
and it may become an important tool to study planet formation processes.

\section*{Acknowledgements}
We thank Hidekazu Tanaka for providing the detailed results of his past papers to us.
S.I. acknowledges the financial support of JSPS Kakenhi 15H02065 and MEXT Kakenhi 18H05438. 
S.M. thanks STFC (ST/S000399/1) for the financial support. She is also grateful for the hospitality by ELSI during her visit there.
R.B. is grateful for financial assistance from JSPS Shingakujutsu Kobo (JP19H05071). 

\appendix
\section{Derivation of Eqs.~(13) and (14)}

The original formulae by \citet{Papaloizou00} are
\begin{align}
\tau_e^{-1} & \simeq \frac{1}{2.5\times 10^3} 
\left(\frac{\epsilon}{0.4H}\right)^{-2.5} 
\left[1 + \frac{1}{4} \hat{e}^{3} \right]^{-1} \nonumber\\
 & \times \left(\frac{h}{0.07}\right)^{-4}
\left(\frac{M_{\rm GD}}{2 M_{\rm J}}\right)
\left(\frac{M_{\rm p}}{M_{\oplus}}\right)
\left(\frac{r}{\rm 1\, au}\right)^{-1} {\rm yr}^{-1}
\label{eq:PL00_te0} \\
\tau_m^{-1} & \simeq \frac{1}{3.5\times 10^5} 
\left(\frac{\epsilon}{0.4H}\right)^{-1.75} 
\left[\frac{1 - \left( \frac{\hat{e}}{1.1} \right)^4} 
{1 + \left( \frac{\hat{e}}{1.3} \right)^5} \right] \nonumber\\
 & \times \left(\frac{h}{0.07}\right)^{-2}
\left(\frac{M_{\rm GD}}{2 M_{\rm J}}\right)
\left(\frac{M_{\rm p}}{M_{\oplus}}\right)
\left(\frac{r}{\rm 1\, au}\right)^{-1} {\rm yr}^{-1}
\label{eq:PL00_tm0}
\end{align}
where $M_{\rm GD}$ is the gas mass within 5 au
in their model with $q=3/2$, 
\begin{align}
M_{\rm GD} & \simeq \int^{\rm 5 \, au} 2 \pi r \Sigma dr
= \int^{\rm 5 \, au} 2 \pi r \Sigma_{\rm 1\,au} \left(\frac{r}{\rm 1\, au}\right)^{-3/2} dr 
\nonumber\\
 & = 4\sqrt{5} \, \pi \Sigma_{\rm 1\,au} ({\rm 1 \, au})^2
= 4\sqrt{5} \, \pi \Sigma r^2 \left(\frac{r}{\rm 1\, au}\right)^{-1/2}. 
\end{align}
For comparison with other formulae, 
we scale these timescales by $t_{\rm wave}$ (Eq.~(\ref{eq:twave})),
\begin{align}
& t_{\rm wave}^{-1} = \left(\frac{M_{\rm p}}{M_*}\right)\left(\frac{\Sigma r^2}{M_*}\right)
h^{-4}  \Omega_K.
\end{align}
From this equation, 
\begin{align}
& \left(\frac{h}{0.07}\right)^{-4} 
\left(\frac{M_{\rm GD}}{2 M_{\rm J}}\right)
\left(\frac{M_{\rm p}}{M_{\oplus}}\right)
\left(\frac{r}{\rm 1\, au}\right)^{-1} {\rm yr}^{-1} \nonumber \\
& = 0.07^4 \times 4\sqrt{5} \, \pi 
\left(\frac{M_*}{2M_{\rm J}}\right)
\left(\frac{\Sigma r^2}{M_*}\right) \nonumber \\
& \hspace*{0.5cm} \times 
\left(\frac{M_*}{M_\oplus}\right)
\left(\frac{M_{\rm p}}{M_*}\right)
\left(\frac{r}{\rm 1\, au}\right)^{-3/2} {\rm yr}^{-1} \nonumber \\
& = 0.07^4 \times 2 \sqrt{5} \times
524 \times (3.33 \times 10^5) \left(\frac{M_*}{M_\odot}\right)^{3/2}\, t_{\rm wave}^{-1}\nonumber \\
& \simeq 1.86 \times 10^4  \left(\frac{M_*}{M_\odot}\right)^{3/2}\, t_{\rm wave}^{-1}.
\end{align}
Substituting this relation into Eqs.~(\ref{eq:PL00_te0}) and (\ref{eq:PL00_tm0})
adopting the $\epsilon$ by $0.5H$, we obtain
\begin{align}
\tau_e^{-1} & \simeq  4.26 \, \left(\frac{\epsilon}{0.5H}\right)^{-2.5} t_{\rm wave}^{-1} \left[1 + \frac{1}{4} \hat{e}^{3} \right]^{-1},
\label{eq:PL00_te1} \\
\tau_m^{-1} & \simeq 7.33 \, \left(\frac{\epsilon}{0.5H}\right)^{-1.75} h^{2} \, t_{\rm wave}^{-1} 
\frac{1 - \left( \frac{\hat{e}}{1.1} \right)^4} 
{1 + \left( \frac{\hat{e}}{1.3} \right)^5}.
\label{eq:PL00_tm1}
\end{align}

\section{Chandrasekhar's dynamical friction formula}

The well-known Chandrasekhar's dynamical friction formula for stellar dynamics is
\citep{Chandrasekhar43},
\begin{align}
\vct{F}_{\rm DF,Ch} \simeq - 4\pi \ln \Lambda \frac{G^2 \rho M_p}{(\Delta v)^3}
\left[{\rm erf}(X) - \frac{2X}{\sqrt{\pi}}e^{-X^2}\right] \Delta \vct{v},
\end{align}
where $X=\Delta v/\sqrt{2}\sigma$, $\sigma$ is
the velocity dispersion of the field stars with a Maxwellian velocity distribution,
and $\ln \Lambda$ is the 3D log-divergence term, which cannot be a large value in a disc with finite thickness.
Using $\rho = \Sigma/(\sqrt{2\pi} \sigma/\Omega)$,
$G^2 = \Omega^4 r^6/M_*^2$, 
${\rm erf}(X)\sim 1$ for $X \ga 2$, and
$[{\rm erf}(X) - 2X e^{-X^2} /\sqrt{\pi}]
\sim (2/\sqrt{\pi})[X - X^3/3 - X(1-X^2)] 
\sim (4X^3/3\sqrt{\pi})$
for $X < 1$, the subsonic and supersonic limits are
\begin{equation}
\vct{F}_{\rm DF,Ch} \simeq
\left\{ \begin{array}{ll}
\displaystyle 
 - \frac{4}{3} \ln(\Lambda) \, \frac{\Delta \vct{v}}{t_{\rm df}} 
& [{\rm for}\; \Delta v \ll \sigma] 
 \\
\displaystyle 
 - 2 \sqrt{2\pi} \ln(\Lambda)\, \frac{\Delta \vct{v}}{t_{\rm df}}\left( \frac{\Delta v}{\sigma} \right)^{-3}
 & [{\rm for}\; \Delta v \gg \sigma],
\end{array}
\right.
\label{eq:FDF_Ch}
\end{equation}
where 
\begin{equation}
t_{\rm df}^{-1} = \left(\frac{M_p}{M_*}\right)\left(\frac{\Sigma r^2}{M_*}\right)
\left(\frac{\sigma}{v_{K}}\right)^{-4} \Omega_K.
\end{equation}
The $\ln(\Lambda)$ term for planetesimals is $\sim O(1)$
\citep{Ohtsuki02}. 
The velocity dispersion of the field stars in this formula can be
identified as the sound velocity of planet-disc interactions.
In this case, Eq.~(\ref{eq:FDF_Ch}) agrees with
Eq.~(\ref{eq:FDF_Muto}) in both supersonic and subsonic regimes,
except a small difference in the numerical factor,
if $\tilde{\epsilon} = \epsilon/0.5H$ and $\alpha$ are $\sim O(1)$ in Eq.~(\ref{eq:FDF_Muto}). 

\section{Effects of non-isothermal migration}

Because type I migration in the subsonic regime is caused by a residual between
the torque exerted from the inner disc and that from the outer disc, the migration speed and 
even the migration direction (inward or outward) are affected by the disc structure 
\citep{Paardekooper11}.

Following the prescription by \citet{Coleman14},
we incorporate the finite eccentricity correction factors into $\tau_e^{-1}$ and the Lindblad parts of $\tau_m^{-1}$ and $\tau_a^{-1}$.
The corotation part does not exist in the supersonic regime, so we follow 
\citet{Fendyke14} and introduce 
the factor of $\exp(-e/e_f)$ where $e_f = 0.01 + h/2$
for the corotation part. 
The formulae for the non-isothermal disc, instead of 
Eqs.~(\ref{eq:taue_f2}), and (\ref{eq:taua_f2}), and (\ref{eq:taum_f2}), are:
\begin{align}
\tau_e^{-1} & \simeq 0.780 \, t_{\rm wave}^{-1}
\left(1 + \frac{1}{15} \hat{e}^{3} \right)^{-1}, \label{eq:non_iso_e}\\
\tau_a^{-1} & \simeq 2 h^{2} t_{\rm wave}^{-1} 
\left[ C_{\rm P}  \left( 1 + \frac{C_{\rm P}}{C_{\rm M}} \hat{e} \right)^{-1}
- \frac{\Gamma_C}{\Gamma_0} \exp \left( -\frac{e}{e_f} \right) \right], \label{eq:non_iso_a}\\
\tau_m^{-1} & \simeq \frac{1}{2}\tau_a^{-1} - e^2 \tau_e^{-1} \nonumber \\
 & \simeq h^{2} t_{\rm wave}^{-1} 
 \left[ \frac{C_{\rm P}}{1 + \frac{C_{\rm P}}{C_{\rm M}} \hat{e}} 
 - \frac{\Gamma_C}{\Gamma_0} \exp \left( -\frac{e}{e_f} \right)
 - \frac{ 0.780 \, \hat{e}^2 }{ 1 + \frac{1}{15} \hat{e}^3 } 
 \right],
\label{eq:non_iso_m}
\end{align}
where $\Gamma_C$ is the corotation torque and $C_{\rm P} = 2.5 + 1.7 q - 0.1 p$.
Detailed form of the scaled corotation torque, $\Gamma_C/\Gamma_0$, 
is described by \citet{Paardekooper11}.

\section{Inclination damping}

In this paper, we neglect orbital inclinations, assuming that $i$ is much smaller than $e$,
although we considered three dimensional interactions.
However, in a similar way to the $e$-damping formula, the $i$-damping formula can be derived, as long as $i < h$.
For $i > h$, the planetary orbit is not within the disc in most of time and
the disc-planet interactions are weak \citep{Rein12}.
Here we consider the case of $i < h$.
 
The inclinations also contributed to the relative velocity, while
the contribution is less than that by $e$.
The formulae with $i$ should be as follows:
\begin{align}
\tau_e^{-1} & \simeq 0.780 \, t_{\rm wave}^{-1}
\left[1 + \frac{1}{15} (\hat{e}^2+\hat{i}^2)^{3/2} \right]^{-1}, \label{eq:non_iso_e2}\\
\tau_i^{-1} & \simeq 0.544 \, t_{\rm wave}^{-1}
\left[1 + \frac{1}{21.5} (\hat{e}^2+\hat{i}^2)^{3/2} \right]^{-1}, \label{eq:non_iso_e2}\\
\tau_m^{-1} & \simeq \frac{1}{2}\tau_a^{-1} - e^2 \tau_e^{-1} 
- i^2\tau_{i}^{-1}. 
 \label{eq:non_iso_m2}
\end{align}
The migration rate for the isothermal case is
\begin{align}
\tau_a^{-1} & \simeq C_{\rm T} h^{2} 
\left( 1 + \frac{C_{\rm T}}{C_{\rm M}} (\hat{e}^2+\hat{i}^2)^{1/2}\right)^{-1}t_{\rm wave}^{-1},
\end{align}
and for the non-isothermal case,
\begin{align}
\tau_a^{-1} & \simeq 2 h^{2} t_{\rm wave}^{-1}  \nonumber \\
\times & \left[ C_{\rm P}  \left( 1 + \frac{C_{\rm P}}{C_{\rm M}} \sqrt{\hat{e}^2+\hat{i}^2} \right)^{-1}
- \frac{\Gamma_C}{\Gamma_0} \exp \left( -\frac{\sqrt{e^2+i^2}}{e_f} \right) \right].\label{eq:non_iso_a2}
\end{align}







\bsp	
\label{lastpage}

\section*{References}
\begin{thebibliography}{}
\makeatletter
\relax
\def\mn@urlcharsother{\let\do\@makeother \do\$\do\&\do\#\do\^\do\_\do\%\do\~}
\def\mn@doi{\begingroup\mn@urlcharsother \@ifnextchar [ {\mn@doi@}
  {\mn@doi@[]}}
\def\mn@doi@[#1]#2{\def\@tempa{#1}\ifx\@tempa\@empty \href
  {http://dx.doi.org/#2} {doi:#2}\else \href {http://dx.doi.org/#2} {#1}\fi
  \endgroup}
\def\mn@eprint#1#2{\mn@eprint@#1:#2::\@nil}
\def\mn@eprint@arXiv#1{\href {http://arxiv.org/abs/#1} {{\tt arXiv:#1}}}
\def\mn@eprint@dblp#1{\href {http://dblp.uni-trier.de/rec/bibtex/#1.xml}
  {dblp:#1}}
\def\mn@eprint@#1:#2:#3:#4\@nil{\def\@tempa {#1}\def\@tempb {#2}\def\@tempc
  {#3}\ifx \@tempc \@empty \let \@tempc \@tempb \let \@tempb \@tempa \fi \ifx
  \@tempb \@empty \def\@tempb {arXiv}\fi \@ifundefined
  {mn@eprint@\@tempb}{\@tempb:\@tempc}{\expandafter \expandafter \csname
  mn@eprint@\@tempb\endcsname \expandafter{\@tempc}}}

\bibitem[\protect\citeauthoryear{{Artymowicz}}{{Artymowicz}}{1993}]{Artymowicz93}
{Artymowicz} P.,  1993, \mn@doi [\apj] {10.1086/173469}, \href
  {https://ui.adsabs.harvard.edu/abs/1993ApJ...419..155A} {419, 155}
  
\bibitem[\protect\citeauthoryear{{Baruteau} et~al.,}{{Baruteau}
  et~al.}{2014}]{Baruteau14}
{Baruteau} C.,  et~al., 2014, in {Beuther} H.,  {Klessen} R.~S.,  {Dullemond}
  C.~P.,   {Henning} T.,  eds, Protostars and Planets VI. p.~667 (\mn@eprint
  {arXiv} {1312.4293}), \mn@doi{10.2458/azu_uapress_9780816531240-ch029}

\bibitem[\protect\citeauthoryear{{Bitsch} \& {Kley}}{{Bitsch} \&
  {Kley}}{2010}]{Bitsch10}
{Bitsch} B.,  {Kley} W.,  2010, \mn@doi [\aap] {10.1051/0004-6361/201014414},
  \href {https://ui.adsabs.harvard.edu/abs/2010A&A...523A..30B} {523, A30}

\bibitem[\protect\citeauthoryear{{Brasser}, {Matsumura}, {Muto}  \&
  {Ida}}{{Brasser} et~al.}{2018}]{Brasser18}
{Brasser} R.,  {Matsumura} S.,  {Muto} T.,   {Ida} S.,  2018, \mn@doi [\apjl]
  {10.3847/2041-8213/aada18}, \href
  {https://ui.adsabs.harvard.edu/abs/2018ApJ...864L...8B} {864, L8}

\bibitem[\protect\citeauthoryear{{Chandrasekhar}}{{Chandrasekhar}}{1943}]{Chan%
drasekhar43}
{Chandrasekhar} S.,  1943, \mn@doi [\apj] {10.1086/144517}, \href
  {https://ui.adsabs.harvard.edu/abs/1943ApJ....97..255C} {97, 255}

\bibitem[\protect\citeauthoryear{{Coleman} \& {Nelson}}{{Coleman} \&
  {Nelson}}{2014}]{Coleman14}
{Coleman} G. A.~L.,  {Nelson} R.~P.,  2014, \mn@doi [\mnras]
  {10.1093/mnras/stu1715}, \href
  {https://ui.adsabs.harvard.edu/abs/2014MNRAS.445..479C} {445, 479}

\bibitem[\protect\citeauthoryear{{Coleman} \& {Nelson}}{{Coleman} \&
  {Nelson}}{2016a}]{Coleman16a}
{Coleman} G. A.~L.,  {Nelson} R.~P.,  2016a, \mn@doi [\mnras]
  {10.1093/mnras/stw1177}, \href
  {https://ui.adsabs.harvard.edu/abs/2016MNRAS.460.2779C} {460, 2779}

\bibitem[\protect\citeauthoryear{{Coleman} \& {Nelson}}{{Coleman} \&
  {Nelson}}{2016b}]{Coleman16b}
{Coleman} G. A.~L.,  {Nelson} R.~P.,  2016b, \mn@doi [\mnras]
  {10.1093/mnras/stw1177}, \href
  {https://ui.adsabs.harvard.edu/abs/2016MNRAS.460.2779C} {460, 2779}

\bibitem[\protect\citeauthoryear{{Cossou}, {Raymond}, {Hersant}  \&
  {Pierens}}{{Cossou} et~al.}{2014}]{Cossou14}
{Cossou} C.,  {Raymond} S.~N.,  {Hersant} F.,   {Pierens} A.,  2014, \mn@doi
  [\aap] {10.1051/0004-6361/201424157}, \href
  {https://ui.adsabs.harvard.edu/abs/2014A&A...569A..56C} {569, A56}

\bibitem[\protect\citeauthoryear{CN08}{}]{Cresswell08}
{Cresswell} P.,  {Nelson} R.~P.,  2008, \mn@doi [\aap]
  {10.1051/0004-6361:20079178}, \href
  {https://ui.adsabs.harvard.edu/abs/2008A&A...482..677C} {482, 677}

\bibitem[\protect\citeauthoryear{{Cresswell}, {Dirksen}, {Kley}  \&
  {Nelson}}{{Cresswell} et~al.}{2007}]{Cresswell07}
{Cresswell} P.,  {Dirksen} G.,  {Kley} W.,   {Nelson} R.~P.,  2007, \mn@doi
  [\aap] {10.1051/0004-6361:20077666}, \href
  {https://ui.adsabs.harvard.edu/abs/2007A&A...473..329C} {473, 329}

\bibitem[\protect\citeauthoryear{{Daisaka}, {Tanaka}  \& {Ida}}{{Daisaka}
  et~al.}{2006}]{Kominami06}
{Daisaka} J.~K.,  {Tanaka} H.,   {Ida} S.,  2006, \mn@doi [\icarus]
  {10.1016/j.icarus.2006.07.003}, \href
  {https://ui.adsabs.harvard.edu/abs/2006Icar..185..492D} {185, 492}

\bibitem[\protect\citeauthoryear{{Fendyke} \& {Nelson}}{{Fendyke} \&
  {Nelson}}{2014}]{Fendyke14}
{Fendyke} S.~M.,  {Nelson} R.~P.,  2014, \mn@doi [\mnras]
  {10.1093/mnras/stt1867}, \href
  {https://ui.adsabs.harvard.edu/abs/2014MNRAS.437...96F} {437, 96}

\bibitem[\protect\citeauthoryear{{Goldreich} \& {Schlichting}}{{Goldreich} \&
  {Schlichting}}{2014}]{Goldreich14}
{Goldreich} P.,  {Schlichting} H.~E.,  2014, \mn@doi [\aj]
  {10.1088/0004-6256/147/2/32}, \href
  {https://ui.adsabs.harvard.edu/abs/2014AJ....147...32G} {147, 32}

\bibitem[\protect\citeauthoryear{{Goldreich} \& {Tremaine}}{{Goldreich} \&
  {Tremaine}}{1979}]{Goldreich79}
{Goldreich} P.,  {Tremaine} S.,  1979, \mn@doi [\apj] {10.1086/157448}, \href
  {https://ui.adsabs.harvard.edu/abs/1979ApJ...233..857G} {233, 857}
  
\bibitem[\protect\citeauthoryear{{Grishin} \& {Perets}}{{Grishin} \&
  {Perets}}{2015}]{Grishin15}
{Grishin} E.,  {Perets} H.~B.,  2015, \mn@doi [\apj]
  {10.1088/0004-637X/811/1/54}, \href
  {https://ui.adsabs.harvard.edu/abs/2015ApJ...811...54G} {811, 54}

\bibitem[\protect\citeauthoryear{{Henon} \& {Petit}}{{Henon} \&
  {Petit}}{1986}]{Henon86}
{Henon} M.,  {Petit} J.~M.,  1986, \mn@doi [Celestial Mechanics]
  {10.1007/BF01234287}, \href
  {https://ui.adsabs.harvard.edu/abs/1986CeMec..38...67H} {38, 67}

\bibitem[\protect\citeauthoryear{{Ida} \& {Lin}}{{Ida} \& {Lin}}{2010}]{Ida10}
{Ida} S.,  {Lin} D.~N.~C.,  2010, \mn@doi [\apj] {10.1088/0004-637X/719/1/810},
  \href {https://ui.adsabs.harvard.edu/abs/2010ApJ...719..810I} {719, 810}

\bibitem[\protect\citeauthoryear{{Ida}, {Guillot}  \& {Morbidelli}}{{Ida}
  et~al.}{2016}]{Ida16}
{Ida} S.,  {Guillot} T.,   {Morbidelli} A.,  2016, \mn@doi [\aap]
  {10.1051/0004-6361/201628099}, \href
  {https://ui.adsabs.harvard.edu/abs/2016A&A...591A..72I} {591, A72}

\bibitem[\protect\citeauthoryear{{Izidoro}, {Bitsch}, {Raymond}, {Johansen},
  {Morbidelli}, {Lambrechts}  \& {Jacobson}}{{Izidoro}
  et~al.}{2019}]{Izidoro19}
{Izidoro} A.,  {Bitsch} B.,  {Raymond} S.~N.,  {Johansen} A.,  {Morbidelli} A.,
   {Lambrechts} M.,   {Jacobson} S.~A.,  2019, arXiv e-prints, \href
  {https://ui.adsabs.harvard.edu/abs/2019arXiv190208772I} {p. arXiv:1902.08772}

\bibitem[\protect\citeauthoryear{{Kominami}, {Tanaka}  \& {Ida}}{{Kominami}
  et~al.}{2005}]{Kominami05}
{Kominami} J.,  {Tanaka} H.,   {Ida} S.,  2005, \mn@doi [\icarus]
  {10.1016/j.icarus.2005.05.008}, \href
  {https://ui.adsabs.harvard.edu/abs/2005Icar..178..540K} {178, 540}

\bibitem[\protect\citeauthoryear{{Matsumura}, {Brasser}  \& {Ida}}{{Matsumura}
  et~al.}{2017}]{Matsumura17}
{Matsumura} S.,  {Brasser} R.,   {Ida} S.,  2017, \mn@doi [\aap]
  {10.1051/0004-6361/201731155}, \href
  {https://ui.adsabs.harvard.edu/abs/2017A&A...607A..67M} {607, A67}

\bibitem[\protect\citeauthoryear{{Miyoshi}, {Takeuchi}, {Tanaka}  \&
  {Ida}}{{Miyoshi} et~al.}{1999}]{Miyoshi99}
{Miyoshi} K.,  {Takeuchi} T.,  {Tanaka} H.,   {Ida} S.,  1999, \mn@doi [\apj]
  {10.1086/307086}, \href
  {https://ui.adsabs.harvard.edu/abs/1999ApJ...516..451M} {516, 451}

\bibitem[\protect\citeauthoryear{{Morbidelli} \& {Raymond}}{{Morbidelli} \&
  {Raymond}}{2016}]{Morbidelli16}
{Morbidelli} A.,  {Raymond} S.~N.,  2016, \mn@doi [Journal of Geophysical
  Research (Planets)] {10.1002/2016JE005088}, \href
  {https://ui.adsabs.harvard.edu/abs/2016JGRE..121.1962M} {121, 1962}

\bibitem[\protect\citeauthoryear{MTI11}{}]{Muto11}
{Muto} T.,  {Takeuchi} T.,   {Ida} S.,  2011, \mn@doi [\apj]
  {10.1088/0004-637X/737/1/37}, \href
  {https://ui.adsabs.harvard.edu/abs/2011ApJ...737...37M} {737, 37}

\bibitem[\protect\citeauthoryear{{Ogihara} \& {Ida}}{{Ogihara} \&
  {Ida}}{2009}]{Ogihara09}
{Ogihara} M.,  {Ida} S.,  2009, \mn@doi [\apj] {10.1088/0004-637X/699/1/824},
  \href {https://ui.adsabs.harvard.edu/abs/2009ApJ...699..824O} {699, 824}

\bibitem[\protect\citeauthoryear{{Ogihara}, {Ida}  \& {Morbidelli}}{{Ogihara}
  et~al.}{2007}]{Ogihara07}
{Ogihara} M.,  {Ida} S.,   {Morbidelli} A.,  2007, \mn@doi [\icarus]
  {10.1016/j.icarus.2006.12.006}, \href
  {https://ui.adsabs.harvard.edu/abs/2007Icar..188..522O} {188, 522}
  
  \bibitem[\protect\citeauthoryear{{Ohtsuki}, {Stewart}  \& {Ida}}{{Ohtsuki}
  et~al.}{2002}]{Ohtsuki02}
{Ohtsuki} K.,  {Stewart} G.~R.,   {Ida} S.,  2002, \mn@doi [Icarus]
  {10.1006/icar.2001.6741}, \href
  {https://ui.adsabs.harvard.edu/abs/2002Icar..155..436O} {155, 436}

\bibitem[\protect\citeauthoryear{{Ostriker}}{{Ostriker}}{1999}]{Ostriker99}
{Ostriker} E.~C.,  1999, \mn@doi [\apj] {10.1086/306858}, \href
  {https://ui.adsabs.harvard.edu/abs/1999ApJ...513..252O} {513, 252}

\bibitem[\protect\citeauthoryear{{Paardekooper}, {Baruteau}  \&
  {Kley}}{{Paardekooper} et~al.}{2011}]{Paardekooper11}
{Paardekooper} S.~J.,  {Baruteau} C.,   {Kley} W.,  2011, \mn@doi [\mnras]
  {10.1111/j.1365-2966.2010.17442.x}, \href
  {https://ui.adsabs.harvard.edu/abs/2011MNRAS.410..293P} {410, 293}

\bibitem[\protect\citeauthoryear{PL00}{}]{Papaloizou00}
{Papaloizou} J.~C.~B.,  {Larwood} J.~D.,  2000, \mn@doi [\mnras]
  {10.1046/j.1365-8711.2000.03466.x}, \href
  {https://ui.adsabs.harvard.edu/abs/2000MNRAS.315..823P} {315, 823}

\bibitem[\protect\citeauthoryear{{Rein}}{{Rein}}{2012}]{Rein12}
{Rein} H.,  2012, \mn@doi [MNRAS] {10.1111/j.1365-2966.2012.20869.x}, \href
  {https://ui.adsabs.harvard.edu/abs/2012MNRAS.422.3611R} {422, 3611}

\bibitem[\protect\citeauthoryear{{Rephaeli} \& {Salpeter}}{{Rephaeli} \&
  {Salpeter}}{1980}]{Rephaeli80}
{Rephaeli} Y.,  {Salpeter} E.~E.,  1980, \mn@doi [\apj] {10.1086/158202}, \href
  {https://ui.adsabs.harvard.edu/abs/1980ApJ...240...20R} {240, 20}

\bibitem[\protect\citeauthoryear{{Sanchez-Salcedo}}{{Sanchez-Salcedo}}{2019}]{
Sanchez19}
{Sanchez-Salcedo} F.~J.,  2019, \mn@doi [\apj] {10.1086/15820210.3847/1538-4357/ab46ae}, \href
  {https://ui.adsabs.harvard.edu/abs/2019arXiv191003024Shttps://ui.adsabs.harvard.edu/abs/2019ApJ...885..152S} {885, 152}

\bibitem[\protect\citeauthoryear{TW04}{}]{Tanaka04}
{Tanaka} H.,  {Ward} W.~R.,  2004, \mn@doi [\apj] {10.1086/380992}, \href
  {https://ui.adsabs.harvard.edu/abs/2004ApJ...602..388T} {602, 388}

\bibitem[\protect\citeauthoryear{{Tanaka}, {Takeuchi}  \& {Ward}}{{Tanaka}
  et~al.}{2002}]{Tanaka02}
{Tanaka} H.,  {Takeuchi} T.,   {Ward} W.~R.,  2002, \mn@doi [\apj]
  {10.1086/324713}, \href
  {https://ui.adsabs.harvard.edu/abs/2002ApJ...565.1257T} {565, 1257}

\bibitem[\protect\citeauthoryear{{Terquem} \& {Papaloizou}}{{Terquem} \&
  {Papaloizou}}{2007}]{Terquem07}
{Terquem} C.,  {Papaloizou} J. C.~B.,  2007, \mn@doi [\apj] {10.1086/509497},
  \href {https://ui.adsabs.harvard.edu/abs/2007ApJ...654.1110T} {654, 1110}

\bibitem[\protect\citeauthoryear{{Vicente}, {Cardoso}  \&
  {Zilh{\~a}o}}{{Vicente} et~al.}{2019}]{Vincente19}
{Vicente} R.,  {Cardoso} V.,   {Zilh{\~a}o} M.,  2019, \mn@doi [\mnras]
  {10.1093/mnras/stz2526}, \href
  {https://ui.adsabs.harvard.edu/abs/2019MNRAS.489.5424V} {489, 5424}
  
  \bibitem[\protect\citeauthoryear{{Ward}}{{Ward}}{1997}]{Ward97}
{Ward} W.~R.,  1997, \mn@doi [Icarus] {10.1006/icar.1996.5647}, \href
  {https://ui.adsabs.harvard.edu/abs/1997Icar..126..261W} {126, 261}

\makeatother
\end{thebibliography}
\end{document}